\documentclass[sigconf,screen,colorlinks=true,linkcolor=blue,urlcolor=blue,citecolor=blue]{acmart}

\AtBeginDocument{%
  }

\usepackage{subcaption}

\usepackage[colorinlistoftodos,prependcaption,textsize=small]{todonotes}

\setcopyright{acmlicensed}
\copyrightyear{2025}
\acmYear{2025}
\acmDOI{XXXXXXX.XXXXXXX}

\acmConference[CUI '25]{ACM Conference on Conversational User Interfaces}{July 8th–10th, 2025}{Waterloo, Ontario, Canada.}

\acmISBN{978-1-4503-XXXX-X/2018/06}

\begin{document}

\title{Multi-Tool Analysis of User Interface \& Accessibility in Deployed Web-Based Chatbots}

\author{Mukesh Rajmohan}
\email{mrajmoha@gmu.edu}
\orcid{0009-0004-8859-2887}
\affiliation{%
  \institution{George Mason University}
  \streetaddress{4400 University Drive}
  \city{Fairfax}
  \state{Virginia}
  \country{USA}
  \postcode{22030}}

  \author{Smit Desai}
\email{sm.desai@northeastern.edu}
\orcid{0000-0001-6983-8838}
\affiliation{%
  \institution{Northeastern University}
  \city{Boston}
  \state{Massachusetts}
  \country{USA}
  }

\author{Sanchari Das}
\email{sdas35@gmu.edu}
\orcid{0000-0003-1299-7867}
\affiliation{%
  \institution{George Mason University}
  \streetaddress{4400 University Drive}
  \city{Fairfax}
  \state{Virginia}
  \country{USA}
  \postcode{22030}}

\renewcommand{\shortauthors}{Rajmohan et al.}

\begin{abstract}
In this work, we present a multi-tool evaluation of $106$ deployed web-based chatbots, across domains like healthcare, education and customer service, comprising both standalone applications and embedded widgets using automated tools (Google Lighthouse, PageSpeed Insights, SiteImprove Accessibility Checker) and manual audits (Microsoft Accessibility Insights). Our analysis reveals that over 80\% of chatbots exhibit at least one critical accessibility issue, and 45\% suffer from missing semantic structures or ARIA role misuse. Furthermore, we found that accessibility scores correlate strongly across tools (e.g., Lighthouse vs PageSpeed Insights, $r = 0.861$), but performance scores do not ($r = 0.436$), underscoring the value of a multi-tool approach. We offer a replicable evaluation insights and actionable recommendations to support the development of user-friendly conversational interfaces.
\end{abstract}



\maketitle
\section{Introduction}
Conversational agents (CAs) have become integral to digital experiences, enabling natural language interaction with systems across domains such as customer service, education, healthcare, and e-commerce~\cite{Eleni_Ilias,tazi2024we,Wang_Lu,Zheng_tang,surani2022understanding}. As an umbrella category, CAs encompass a range of dialog-based systems, including voice assistants, embodied agents, and chatbots. In this work, we focus specifically on chatbots which are defined as text-based conversational agents that interact with users via web interfaces, either as standalone applications or embedded widgets within broader websites~\cite{Cui_wei,Brandtzaeg_Petter}. These web-based chatbots have redefined how users engage with services online, offering immediate, dialogue-driven access to information and support.

As natural language processing (NLP) and dialogue systems continue to evolve, users increasingly expect chatbot interactions to be seamless, responsive, and accessible. While much of the research and development has focused on improving conversational accuracy and backend intelligence, comparatively less attention has been given to the front-end user experience especially regarding accessibility~\cite{Lister_Kate}. For users relying on assistive technologies such as screen readers, voice navigation, or keyboard-only access, the design and structure of a chatbot interface directly impacts the ability to perceive, navigate, and engage in functional dialogue~\cite{Moraes_Desid}. Inaccessible design choices can thus impede meaningful interactions, such as understanding chatbot responses, locating input fields, or initiating a conversation, leading to exclusion of users with disabilities~\cite{Mekler_Kasper}.

The Web Content Accessibility Guidelines (WCAG) provide a robust framework for making digital content perceivable, operable, understandable, and robust~\cite{wcag21}. Despite the availability of these standards, many web-based chatbot implementations fail to meet even baseline accessibility criteria due to inconsistent markup, poor semantic structure, or limited developer awareness~\cite{Moraes_Desid}. This lack of compliance risks alienating a significant portion of users, undermining the inclusive potential of conversational interfaces.

To address this gap, we conduct a systematic evaluation of $106$ deployed chatbot interfaces, encompassing both standalone chatbot applications (dedicated pages solely for chatbot interaction) and embedded chatbot widgets (integrated components within larger web pages). Including both types allows us to capture the diversity of real-world deployments and assess whether interface context affects accessibility outcomes. Our analysis combines automated assessment tools including: Google Lighthouse, PageSpeed Insights, and SiteImprove Accessibility Checker with manual evaluation using Microsoft Accessibility Insights. This hybrid approach allows us to examine not just technical violations, but also user-facing accessibility barriers that may be missed by automated scanners.

The primary \textbf{contribution} of this work lies in the empirical insights derived from a large-scale, multi-tool evaluation of $106$ deployed web-based chatbot interfaces, encompassing both standalone applications and embedded widgets. Our analysis reveals critical patterns of accessibility noncompliance, deviations from UI development best practices, and structural deficiencies that directly impair inclusive user interaction. We employ an evaluation pipeline incorporating both automated auditing (Google Lighthouse, PageSpeed Insights, SiteImprove Accessibility Checker) and a manual inspection tool (Microsoft Accessibility Insights). This hybrid methodology facilitates a multi-layered diagnostic analysis across technical accessibility metrics. Automated tools enable scalable detection of low-level violations such as missing ARIA roles, improper focus management, and insufficient contrast ratios which are not readily apparent to sighted users but critically impact those using assistive technologies. These automated evaluations function as consistent, reproducible mechanisms for benchmarking interface accessibility at scale, uncovering systemic design flaws that would be prohibitively time-intensive to identify through manual inspection alone. Moreover, they serve as proxies for broader usability challenges, supporting accessibility auditing in scenarios where direct user testing is impractical. This dual-mode analysis advances beyond superficial standards compliance to critically assess how interface-level design decisions manifest in measurable accessibility outcomes.

\section{Related Work}
As conversational agents become increasingly integrated into web environments, researchers have sought to understand not only how users interact with these systems but also how their design impacts usability, inclusivity, and overall experience. Researchers have extensively documented the evolution of chatbot interfaces, tracing their development from early text-based systems to contemporary multi-modal agents that incorporate voice, gesture, and visual components~\cite{Caldarini_Jaf_McGarry_2022,kalhor2023evaluating,Cowan_fischer}. This trajectory reflects broader advances in natural language processing and user experience design, which have enabled more naturalistic and context-aware interactions.

Within this trajectory, several frameworks have aimed to define the essential components of effective conversational interfaces. Brandtzæg and Følstad emphasized the role of interaction transparency and timely system feedback in shaping positive user experiences~\cite{Brandtzaeg_Følstad_2017}. Their framework highlighted how cues such as typing indicators, message timing, and conversational turn-taking enhance perceived responsiveness and trust. Iniesto et al. further explored the tension between designing for rich dialogue functionality and maintaining a simple, usable interface, particularly in support and educational contexts~\cite{Iniesto_Coughlan_Lister_Devine_Freear_Greenwood_Holmes_Kenny_McLeod_Tudor_2023}. These contributions offer valuable insights into engagement and design, but they provide limited focus on how such interfaces support users with disabilities or those relying on assistive technologies.

In parallel, accessibility research has developed robust methods for evaluating digital systems against established guidelines, most notably the WCAG~\cite{wcag21}. Power et al. found that while automated tools can detect common WCAG violations, they often fail to identify issues that impact users with visual or motor impairments, such as inaccessible form labels or inconsistent focus states~\cite{Power_Freire_Petrie_Swallow_2012}. Vigo et al. advocated for a hybrid approach that combines automated scans with manual audits to improve the accuracy and coverage of accessibility evaluations~\cite{Vigo_Brown_Conway_2013}. These methods are widely used in the context of static websites, yet remain underutilized for dynamic and embedded systems like chatbots, which often feature interactive content, asynchronous updates, and floating interface elements~\cite{Edu_Jide}. Meanwhile, research on the usability of conversational agents has focused primarily on dialogue quality, error handling, and user perceptions~\cite{Wei_Kim}. Ashktorab et al. investigated how users respond to chatbot failures and the kinds of repair strategies they expect during conversational breakdowns~\cite{Ashktorab_Jain_Liao_Weisz_2019}. Langevin et al. proposed a set of heuristics for evaluating conversational experiences, identifying factors such as clarity, personalization, and contextual continuity as key usability indicators~\cite{Langevin_Lordon_Avrahami_Cowan_Hirsch_Hsieh_2021}. Although these studies contribute to our understanding of functional performance, they do not directly address how users access or interact with the front-end interface, especially when accessibility barriers are present.

Our work contributes to this growing body of literature by focusing on the interface layer of chatbot interaction. Rather than examining conversational logic, UX or backend intelligence, we evaluate the structural and perceptual qualities of deployed chatbot interfaces through the lens of accessibility. Specifically, we apply established accessibility auditing tools and manual testing procedures to assess compliance with WCAG principles and uncover barriers that affect users with diverse access needs. This approach addresses a significant gap in current research by extending accessibility evaluation practices into the realm of conversational user interfaces, an area where dynamic content, embedded widgets, and inconsistent markup often create unique challenges~\cite{Moraes_Desid}. 

\section{Method}
To evaluate the extent to which web-based chatbot interfaces support usability, security, and accessibility, we adopted a structured, multi-phase methodology. Our approach aimed to capture a broad spectrum of chatbot implementations, assess their user interface performance and accessibility compliance, and analyze them using a combination of industry-standard evaluation tools. This section outlines our process for identifying suitable chatbot interfaces, selecting appropriate assessment tools, and conducting a systematic evaluation of each interface.

\begin{figure*}[!ht]
  \centering
  \includegraphics[width=0.9\linewidth]{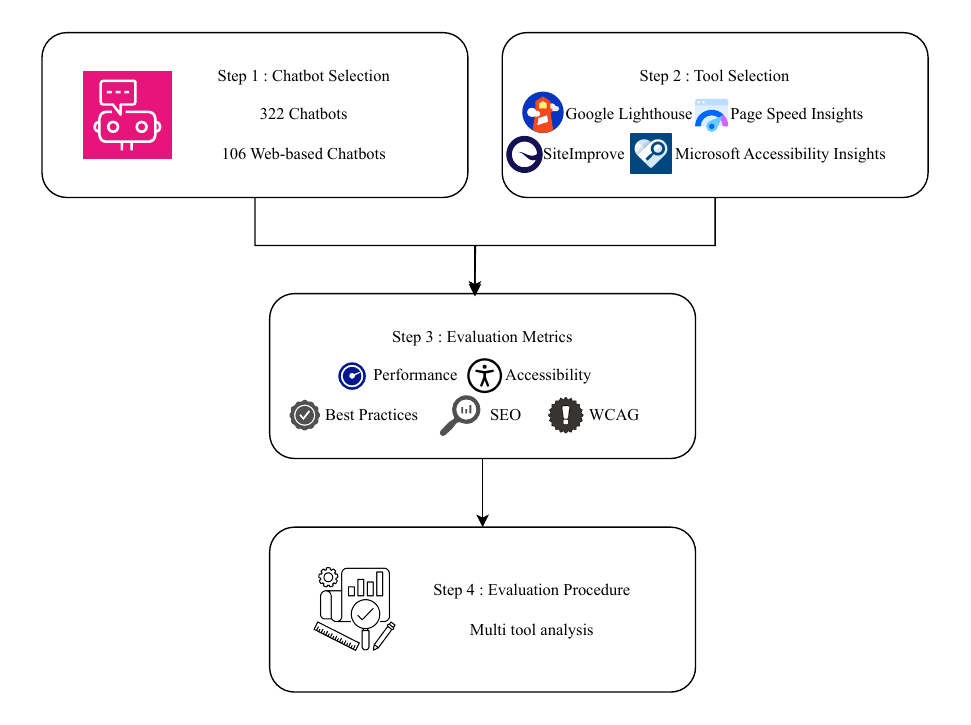}
  \caption{Methodological Overview to Test UI and Accessibility}
  \label{fig:method_diagram}
  \Description{
A four-step flow diagram outlining the methodology for evaluating chatbot interfaces. Step 1 involves selecting 322 chatbots and filtering them down to 106 web-based ones. Step 2 lists the tools used: Google Lighthouse, PageSpeed Insights, SiteImprove Accessibility Checker, and Microsoft Accessibility Insights. Step 3 defines evaluation metrics including performance, accessibility, best practices, SEO, and WCAG violations. Step 4 describes a multi-tool analysis approach integrating both automated and manual evaluations.
}
\end{figure*}

\subsection{Chatbot Selection and Categorization}
We began by constructing a diverse sample of publicly accessible chatbots deployed across various domains. Sources included academic publications, technology blogs, industry directories (e.g., Chatbots.org), and organizational websites identified through targeted search queries such as ``top 10 chatbots in banking'' or ``top 10 chatbots in healthcare'' This process yielded an initial set of $322$ unique chatbots~\footnote{We use chatbots and conversational agents interchangeably in this work} candidates. 

After excluding mobile-only implementations, deprecated links, and non-functional sites through manual validation, we retained a final dataset of $106$ operational web-based chatbots. Our intent was to capture chatbots operating in a variety of sectors, including private enterprises, healthcare services, higher education, and retail, to reflect a broad range of user needs and interaction contexts. Thereafter, we filtered the dataset to retain only those chatbots that were accessible via web browsers. Mobile-exclusive chatbot interfaces were excluded to maintain consistency in evaluation context. We then performed a preliminary screening to ensure that each chatbot remained operational and publicly accessible. Chatbots that had been deprecated, redirected, or failed to load consistently were excluded from further analysis. Following this filtration process, we categorized the remaining chatbots into two interface types based on their deployment model. The first group comprised standalone chatbot applications that functioned as full-page web applications. The second group consisted of embedded chatbot widgets integrated into broader websites, typically accessible via floating buttons or modal pop-ups. 

\subsection{Tool Selection for Evaluation}
To assess each chatbot's interface performance and compliance with accessibility standards, we selected a suite of complementary tools widely used in both academic and industry contexts. The selection aimed to balance automated analysis capabilities with manual validation features, thus enabling a comprehensive evaluation of both surface-level and structural issues in UI design.~\textit{Google Lighthouse}~\cite{lighthouse} and~\textit{PageSpeed Insights}~\cite{pagespeed} were chosen to assess core web vitals, performance optimization, and adherence to best practices, including accessibility and search engine optimization (SEO). Lighthouse provides a standardized auditing protocol for web applications, offering numeric scores across multiple categories. PageSpeed Insights complements this by evaluating load time and rendering efficiency across both mobile and desktop environments.

To evaluate accessibility compliance more specifically, we employed two additional tools:~\textit{SiteImprove Accessibility Checker}~\cite{siteimprove}and~\textit{Microsoft Accessibility Insights}\cite{accessibilityinsights}. SiteImprove facilitates rule-based detection of accessibility issues based on the WCAG, surfacing specific violations such as improper labeling, insufficient contrast, or missing semantic structure. Microsoft Accessibility Insights adds value through its ability to perform both automated and guided manual assessments, offering detailed reports on compliance with WCAG 2.1 Level AA. These tools were selected not only for their analytical capabilities but also for their widespread adoption in accessibility and UI testing practices. They have also been employed by Kishnani et al.\cite{kishnani2023assessing} and Tazi et al.\cite{tazi2023accessibility} in their evaluations of mobile application accessibility.

\subsection{Evaluation Procedure}
Each chatbot in the final dataset was subjected to a standardized evaluation protocol using the four selected tools. For Lighthouse and PageSpeed Insights, we accessed the chatbot's publicly available URL and generated diagnostic reports through Chrome DevTools and the official Google interface, respectively. These reports yielded performance scores and flagged areas requiring improvement. Accessibility audits were conducted using SiteImprove and Microsoft Accessibility Insights browser extensions. Each chatbot interface was interactively scanned to detect WCAG violations, ranging from basic HTML errors to critical issues affecting screen reader compatibility or keyboard navigation. Manual tests were included where automated results were ambiguous or incomplete, such as verifying correct focus order or confirming ARIA role semantics. We recorded all evaluation results in a structured dataset, documenting scores for each key category including: performance, best practices, accessibility, and SEO as well as detailed violation counts and descriptions. This dataset enabled us to perform both comparative and correlation-based analyses, facilitating insights into common challenges and overlooked design flaws in chatbot development. Table~\ref{tab:metrics} explains how the metrics provided by the tools are calculated and how these metrics affect the evaluation of chatbots.

\begin{table*}[h]
\centering
\caption{Overview of Evaluation Metrics for Chatbot Interfaces}
\label{tab:metrics}
\begin{tabular}{|p{3cm}|p{3cm}|p{4cm}|p{3.5cm}|}
\hline
\textbf{Metric Category} & \textbf{What It Measures} & \textbf{Example Indicators} & \textbf{Why It Matters} \\
\hline
\textbf{Performance} & Speed, responsiveness, and visual stability during load & FCP, LCP, TBT, CLS & Impacts perceived quality and usability during interaction \\
\hline
\textbf{Accessibility} & Equitable interaction for all users, especially with assistive technologies & ARIA roles, semantic markup, contrast ratios, keyboard navigation & Ensures inclusive access and compliance with WCAG standards \\
\hline
\textbf{Best Practices} & Adherence to modern web development and security standards & HTTPS use, deprecated API avoidance, valid links & Enhances maintainability, reliability, and user trust \\
\hline
\textbf{SEO} & Discoverability and structural clarity for machines and search engines & Meta tags, heading structure, semantic labels & Reflects content clarity and supports accessibility indirectly \\
\hline
\textbf{WCAG Violations} & Direct counts of non-compliance with accessibility guidelines & Missing labels, contrast errors, focus traps & Provides a granular view of implementation flaws \\
\hline
\end{tabular}
\end{table*}

\section{Results}
The evaluation of $106$ web-based chatbots using revealed consistent gaps in user interface design and accessibility practices, despite the growing sophistication of chatbot dialogue systems. 

\subsection{Tool Agreement and Metric Reliability}
\begin{figure}[!ht]
  \centering
  \includegraphics[width=1.0\linewidth]{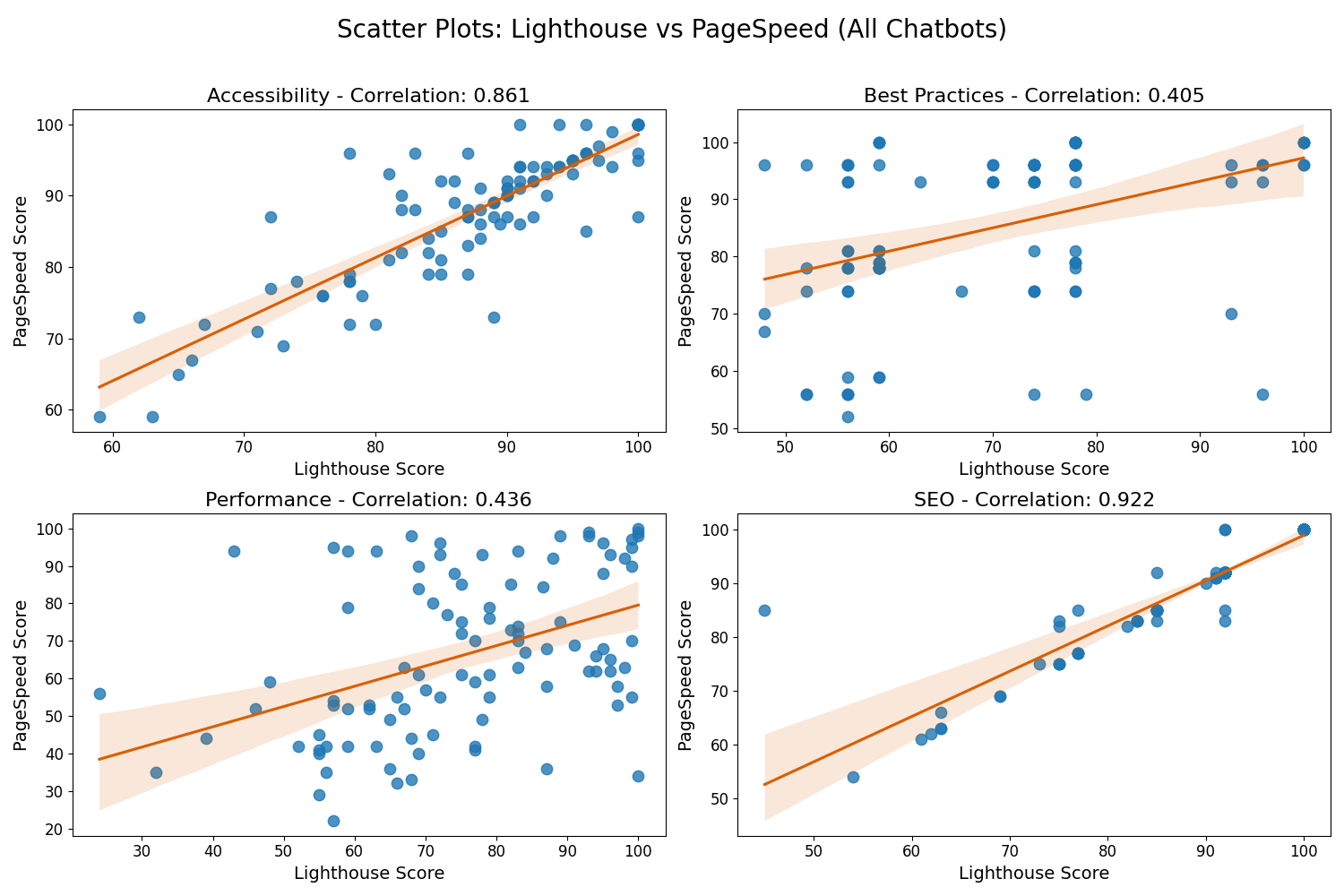}
  \caption{Scatter Plots: Lighthouse vs PageSpeed Insights Metrics (All Chatbots)}
  \Description{
A 2x2 grid of scatter plots comparing scores from Google Lighthouse and PageSpeed Insights across four evaluation metrics: Accessibility, Best Practices, Performance, and SEO. Each plot includes a regression line and a correlation value. Accessibility (0.861) and SEO (0.922) show strong positive correlations, while Performance (0.436) and Best Practices (0.405) exhibit weaker relationships. The data points cluster more tightly around the regression line in the highly correlated plots.
}
  \label{fig:correlation}
\end{figure}

Figure~\ref{fig:correlation} illustrates the correlation between Google Lighthouse and PageSpeed Insights scores across four key categories. Strong positive correlations in Accessibility (r = 0.861) and SEO (r = 0.922) indicate that both tools consistently capture similar issues in these domains. However, Performance (r = 0.436) and Best Practices (r = 0.405) exhibit only weak alignment, highlighting discrepancies in tool focus and scoring methodology. These differences validate our multi-tool approach: no single tool provides a holistic view, and relying solely on one would risk overlooking significant dimensions of chatbot quality, particularly in performance optimization and accessibility nuances.

\subsection{Comparative Performance of Chatbots Across Metrics}

\begin{figure}[!ht]
  \centering
  \includegraphics[width=1.0\linewidth]{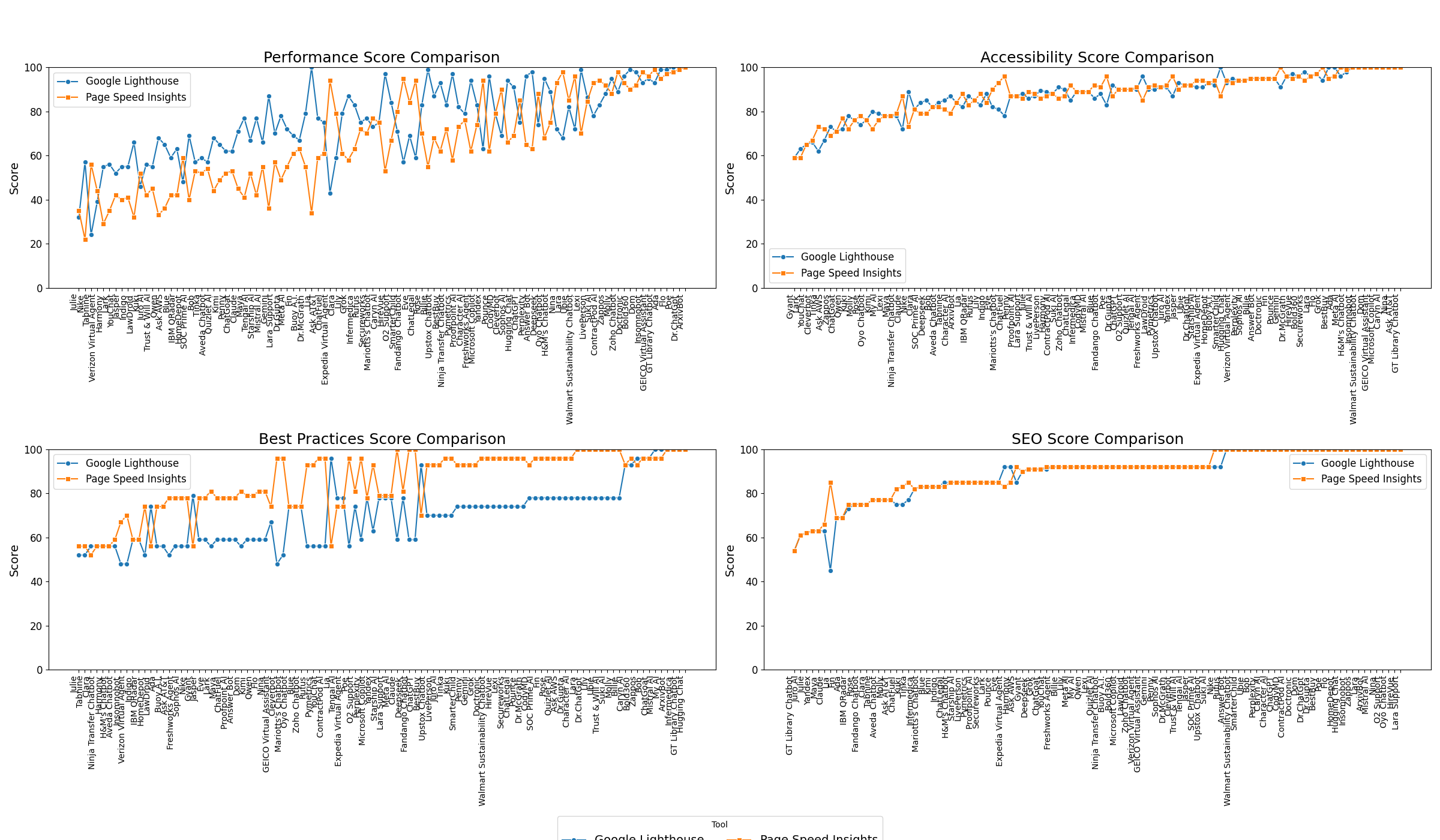}
  \caption{Line Graph: Lighthouse vs PageSpeed Insights Metrics (All Chatbots)}
  \label{fig:line_graph}
  \Description{
A four-panel line chart comparing scores from Google Lighthouse and PageSpeed Insights across 106 chatbot interfaces. Each panel represents a different metric: Performance, Accessibility, Best Practices, and SEO. In each panel, two lines represent Lighthouse and PageSpeed scores for each chatbot. The Accessibility and SEO panels show close alignment between tools, while the Performance and Best Practices panels exhibit more variability and divergence.
}
\end{figure}

\begin{figure}[!ht]
  \centering
  \begin{subfigure}[b]{0.48\linewidth}
    \centering
    \includegraphics[width=\linewidth]{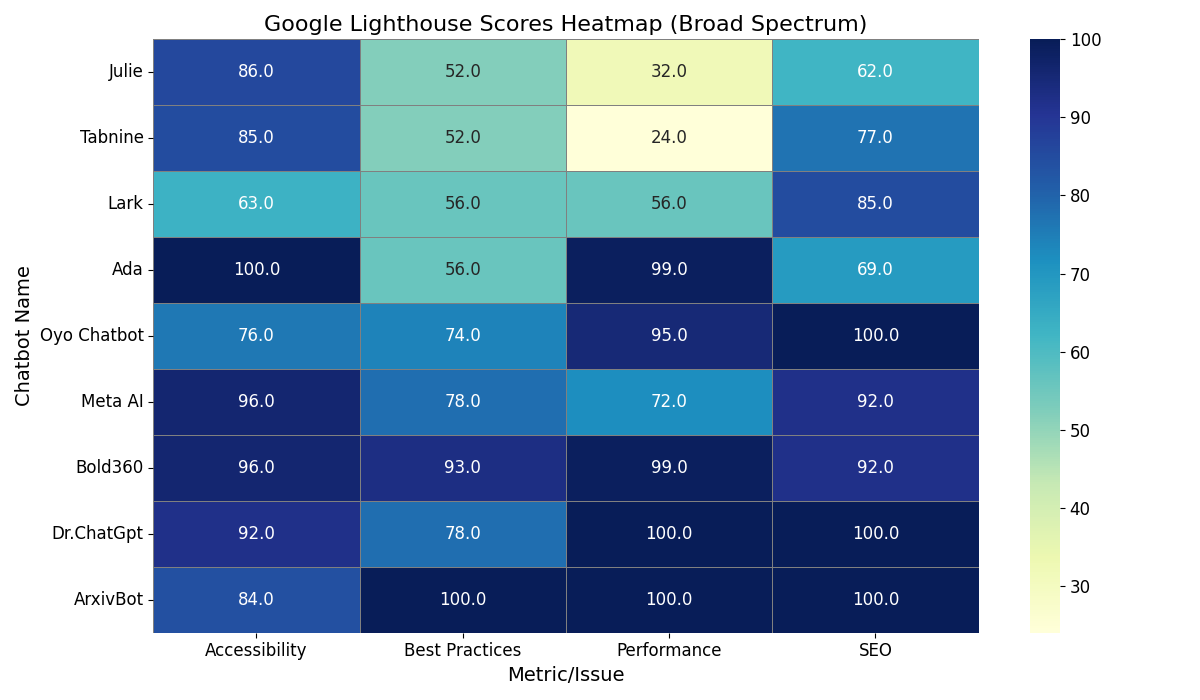}
    \caption{Google Lighthouse Score Heatmap}
    \Description{
A heatmap visualizing Google Lighthouse scores across four categories including, Accessibility, Best Practices, Performance, and SEO for a selected set of chatbots. Rows represent chatbot names and columns show metric scores. Color intensity corresponds to score value, with darker shades indicating higher scores. Dr.ChatGpt, ArxivBot, and Bold360 display consistently high scores across all metrics, while Julie and Tabnine show lower performance, particularly in the Performance category.
}
    \label{fig:heatmap_gl}
  \end{subfigure}
  \hfill
  \begin{subfigure}[b]{0.48\linewidth}
    \centering
    \includegraphics[width=\linewidth]{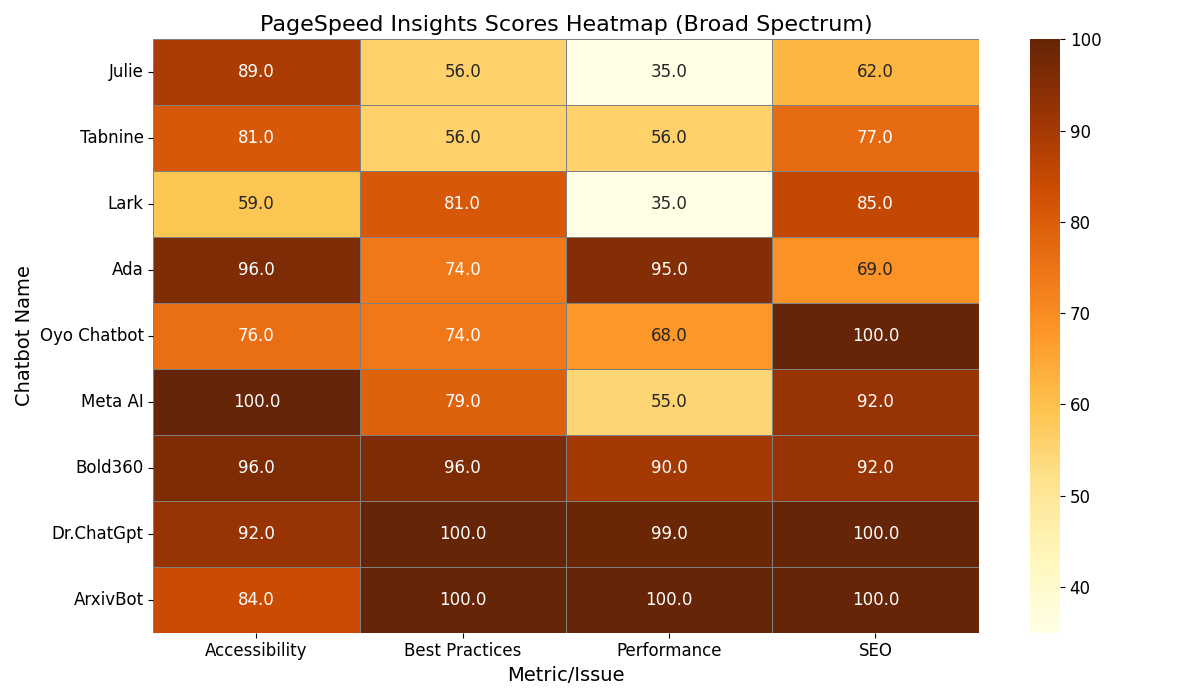}
    \caption{PageSpeed Insights Score Heatmap}
    \label{fig:heatmap_pgi}
  \end{subfigure}
  \caption{Comparison of chatbot performance across evaluation tools. Each heatmap represents metric scores from a different tool.}
  \Description{
A heatmap showing PageSpeed Insights scores for a selected set of chatbots across four categories: Accessibility, Best Practices, Performance, and SEO. Rows represent individual chatbot names, and columns show their respective scores in each category. Color intensity increases with higher scores. Notably, ArxivBot, Dr.ChatGpt, and Bold360 achieve perfect or near-perfect scores across all metrics, while others like Julie and Lark show weaker performance in the Performance category.
}
  \label{fig:heatmap_comparison}
\end{figure}

Figure~\ref{fig:line_graph} presents the overall average score trends across the four evaluation metrics including : Performance ($75.1 - GL, 66.5 - PS$), Best Practices ($70.4 - GL, 85.2 - PS$), SEO ($88.5 - GL, 89.2 - PS$), and Accessibility ($87.4 - GL, 87.8 - PS$) comparing the scores generated by the two tools and highlighting their correlation. Appendix Figure~\ref{fig:avg_score} shows the average scores ($80 - GL, 82 - PS)$) from Google Lighthouse and PageSpeed Insights for all the chatbots. This further supports the earlier analysis, confirming that Accessibility scores are generally lower and vary significantly between the different chatbots.

Figures~\ref{fig:heatmap_gl} and~\ref{fig:heatmap_pgi} show heatmaps from Lighthouse and PageSpeed Insights, respectively, illustrating chatbot-level performance across key evaluation metrics. Due to space constraints, we selected a representative subset of nine chatbots from the total of 106 -- three high-performing, three mid-range, and three low-performing to facilitate a comparative analysis across both tools. Among them, chatbots like~\textit{Dr.ChatGpt},~\textit{Bold360}, and~\textit{ArxivBot} demonstrated consistently strong performance across all metrics. In contrast, others such as~\textit{Julie} and~\textit{Tabnine} exhibited significant performance drop-offs, particularly in adherence to best practices, despite achieving relatively high accessibility scores. This disparity highlights a common tendency among developers to prioritize superficial compliance or conversational capabilities while overlooking deeper UI performance optimizations. The result is often an inconsistent user experience, including persistent accessibility shortcomings.

\subsection{WCAG Violations and Their Impact on UX}
\begin{figure}[!ht]
  \centering
  \begin{subfigure}[b]{0.48\linewidth}
    \centering
    \includegraphics[width=\linewidth]{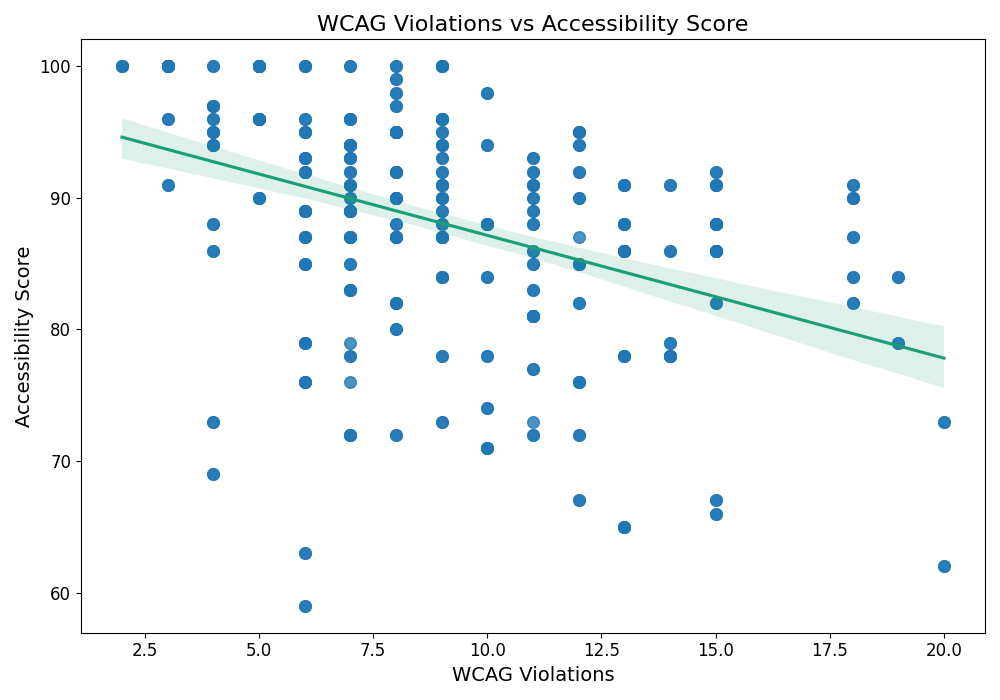}
    \caption{WCAG Violations vs Accessibility Score}
    \Description{A scatter plot showing the relationship between the number of WCAG violations and the accessibility scores of chatbot interfaces. Each dot represents a chatbot. The x-axis shows WCAG violations ranging from 2 to 20, and the y-axis shows accessibility scores from 60 to 100. A clear downward-sloping trendline indicates a strong negative correlation: chatbots with more WCAG violations tend to have significantly lower accessibility scores.}
    \label{fig:wcag_accessibility}
  \end{subfigure}
  \hfill
  \begin{subfigure}[b]{0.48\linewidth}
    \centering
    \includegraphics[width=\linewidth]{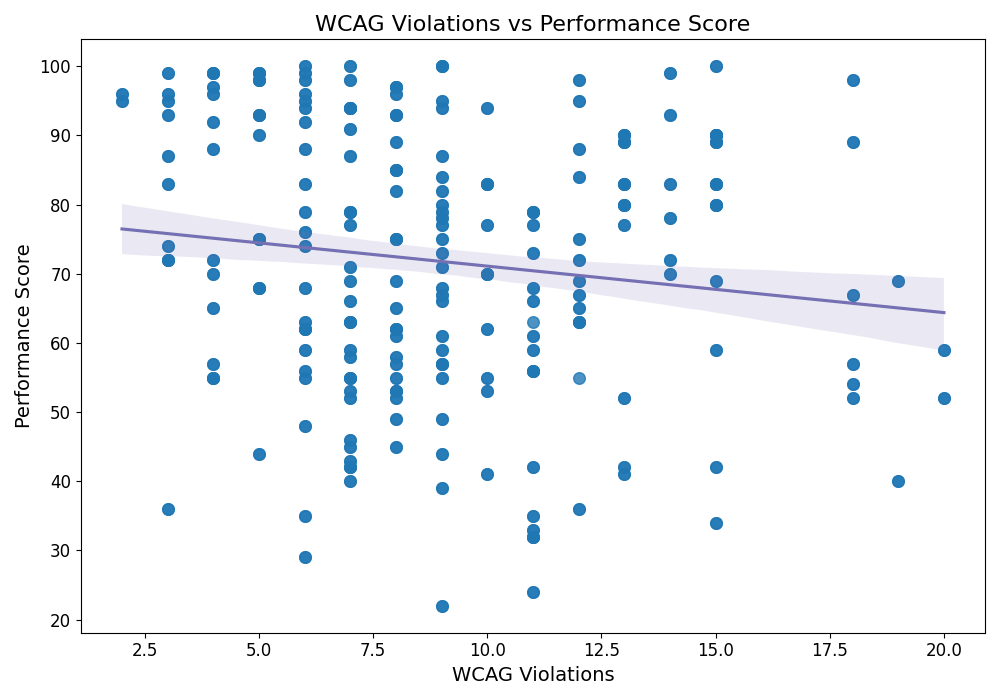}
    \Description{A scatter plot showing the relationship between the number of WCAG violations and the performance scores of chatbot interfaces. Each point represents a chatbot. The x-axis represents WCAG violations ranging from 2 to 20, and the y-axis shows performance scores from 20 to 100. A downward-sloping trendline indicates a weak negative correlation, suggesting that chatbots with more accessibility violations tend to have slightly lower performance scores.}
    \caption{WCAG Violations vs Performance Score}
    \label{fig:wcag_performance}
  \end{subfigure}
  \caption{Relationship between WCAG violations and key evaluation metrics across chatbot interfaces.}
  \label{fig:wcag_comparison}
\end{figure}

Figures~\ref{fig:wcag_accessibility} and~\ref{fig:wcag_performance} explore the relationship between WCAG violation counts and key evaluation metrics. A strong inverse relationship emerges: chatbots with a higher number of violations consistently receive lower Accessibility and Performance scores. This suggests that structural accessibility failures (e.g., missing labels, poor semantic markup) not only impair inclusive interaction but also contribute to degraded performance, such as delayed rendering or broken component loading. These results reinforce that accessibility is not just a legal or ethical requirement, but also a proxy for general UI robustness.

\subsection{Critical Accessibility Barriers}
\begin{figure*}[!ht]
  \centering
  \includegraphics[width=1.1\linewidth]{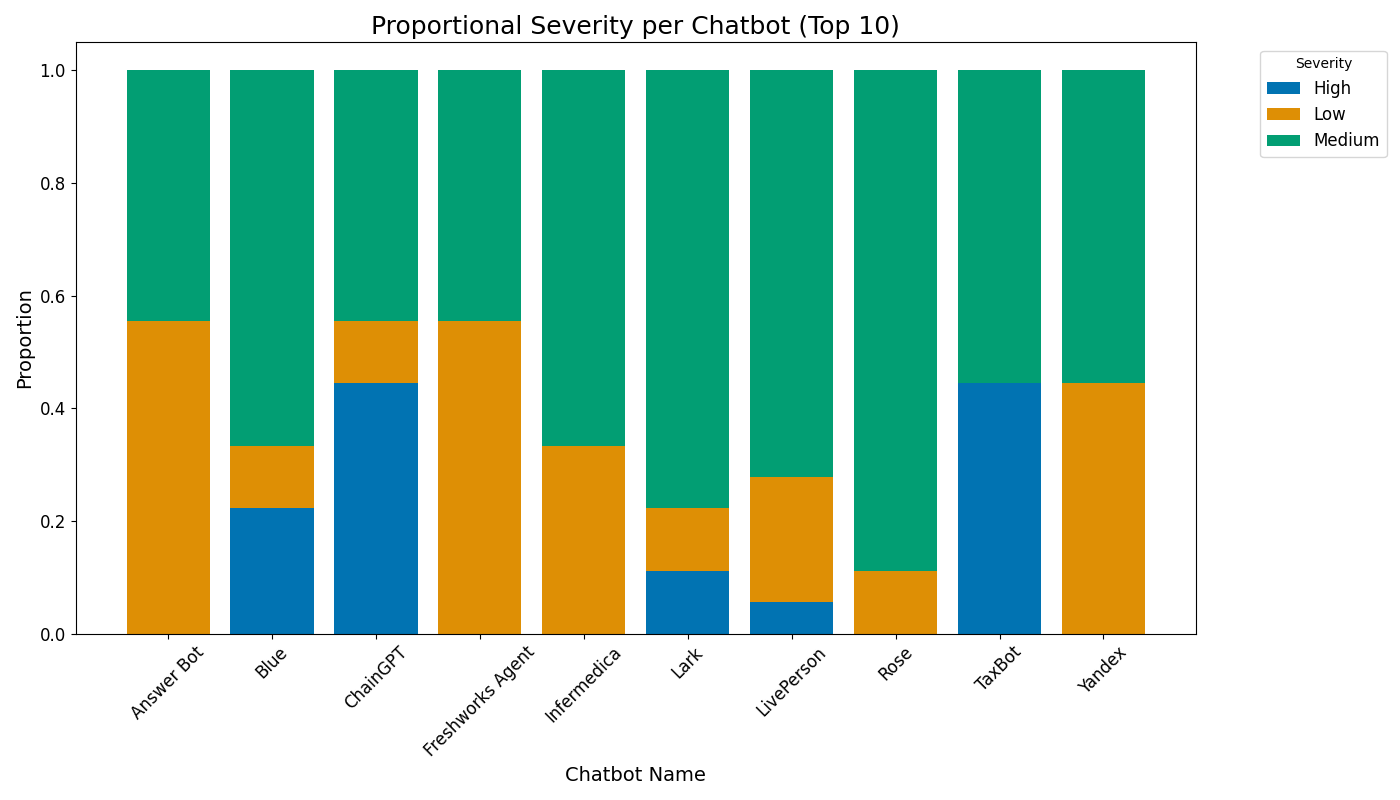}
  \caption{Proportional Severity of Violations by Chatbot (Top 10)}
  \Description{
A stacked bar chart showing the proportional severity of accessibility issues for the top 10 chatbots with the most violations. Each bar is divided into three segments representing high (blue), medium (green), and low (orange) severity issues. Medium-severity issues are the most dominant across chatbots. ChainGPT and TaxBot have higher proportions of high-severity issues, while others like Rose and Lark mostly exhibit medium-severity problems.
}
  \label{fig:severity}
\end{figure*}

Almost 45\% of the issues stem from missing semantic structures or incorrect ARIA role applications core elements for screen reader compatibility. Additionally, the high frequency of visual feedback violations (such as missing focus indicators and keyboard traps) suggests that some chatbots prioritize visual design over accessible interaction. These findings reveal a disconnect between aesthetically driven user interfaces and the functional needs of users who rely on assistive technologies. Figure~\ref{fig:severity} ranks the top $10$ chatbots according to the proportional severity of the accessibility problems detected. While medium-severity violations, such as improper contrast ratios or unclear button roles are prevalent, among the 10 chatbots \textit{ChainGPT} and \textit{TaxBot} show critical violations that can completely block access for users with visual or motor impairments. These include inaccessible modal dialogs, broken tab sequences, or missing focus indicators. Such failures would render the chatbot functionally unusable for some users, underscoring the need for rigorous manual validation in addition to automated audits.

\section{Discussion}
Our findings reveal persistent accessibility and UI shortcomings  across web-based chatbot interfaces. These issues stem not just from technical oversights but from design practices that fail to account for diverse user needs. In this section, we discuss practical design implications to guide the development of more inclusive chatbot interfaces.

\subsection{Prioritize Semantic Integrity in Component Design}
The frequent absence of proper semantic elements and ARIA roles suggests that accessibility is often treated as an implementation detail rather than a core design principle. \textbf{We recommend that developers integrate semantic planning into the early stages of interface design.} This includes assigning appropriate roles to dynamic containers, ensuring that interactive components are programmatically discernible, and maintaining a clear, logical document structure. We advocate that semantic scaffolding is essential not only for screen reader compatibility but also for reliable keyboard navigation and consistent automated testing.

\subsection{Integrate Accessibility into the Visual Design Workflow}
Many chatbots exhibited complex visual design but failed fundamental accessibility checks. This disconnect highlights the need to embed accessibility validation into the visual design pipeline. Designers should evaluate contrast ratios, ensure that icons and controls include descriptive text, and align with accessibility heuristics during prototyping. \textbf{We recommend that collaboration between visual designers and developers is critical and accessibility should not be delegated to implementation alone}. Design systems and component libraries should standardize accessible defaults (e.g., focus indicators, form labels, alt text) to promote consistent inclusion. When visual design is grounded in accessibility principles, the resulting interfaces are not only more inclusive but also more robust and future-proof.

\section{Limitations and Future Work}
Our multi-tool evaluation provides valuable insights into the accessibility of chatbot user interfaces, but it has some limitations. First, we focused exclusively on web-based interfaces accessed through desktop browsers. In future work, we plan to extend our analysis to include app-based chatbots to capture a broader range of interaction contexts. Second, while automated testing tools effectively detect many structural issues, they cannot replicate the lived experiences of users with disabilities. To address this gap, we aim to conduct longitudinal studies that incorporate user testing with individuals who have diverse accessibility needs. 

\section{Conclusion}
As chatbots become more common in digital services, their user interface quality directly affects who can access and use them effectively. While conversational capabilities have advanced, our evaluation of $106$ deployed web-based chatbots reveals widespread shortcomings in accessibility and interface design. Over 80\% of the chatbots we analyzed exhibit critical accessibility issues, including missing ARIA roles, unlabeled buttons, inaccessible modals, and broken keyboard navigation. We found that chatbots with more violations also perform worse in overall metrics: accessibility, performance, best practices, and SEO. Additionally, those with more than ten WCAG violations score 24\% lower on performance metrics related to layout stability and load responsiveness. We also identified distinct failure patterns based on deployment context.These results show that many development teams neglect structural design during chatbot implementation, which undermines usability and excludes users who rely on assistive technologies. 

\begin{acks}
We would like to acknowledge the Center for AI, Privacy, and Security (CAPS) Lab at George Mason University for supporting this work. We would also like to thank the reviewers for their insightful feedback. Any opinions, findings, conclusions, or recommendations expressed in this material are solely those of the authors.
\end{acks}

\bibliographystyle{ACM-Reference-Format}
\bibliography{chatbot}

\appendix
\section{Appendix}

\begin{figure*}[!ht]
  \centering
  \includegraphics[width=1.0\linewidth]{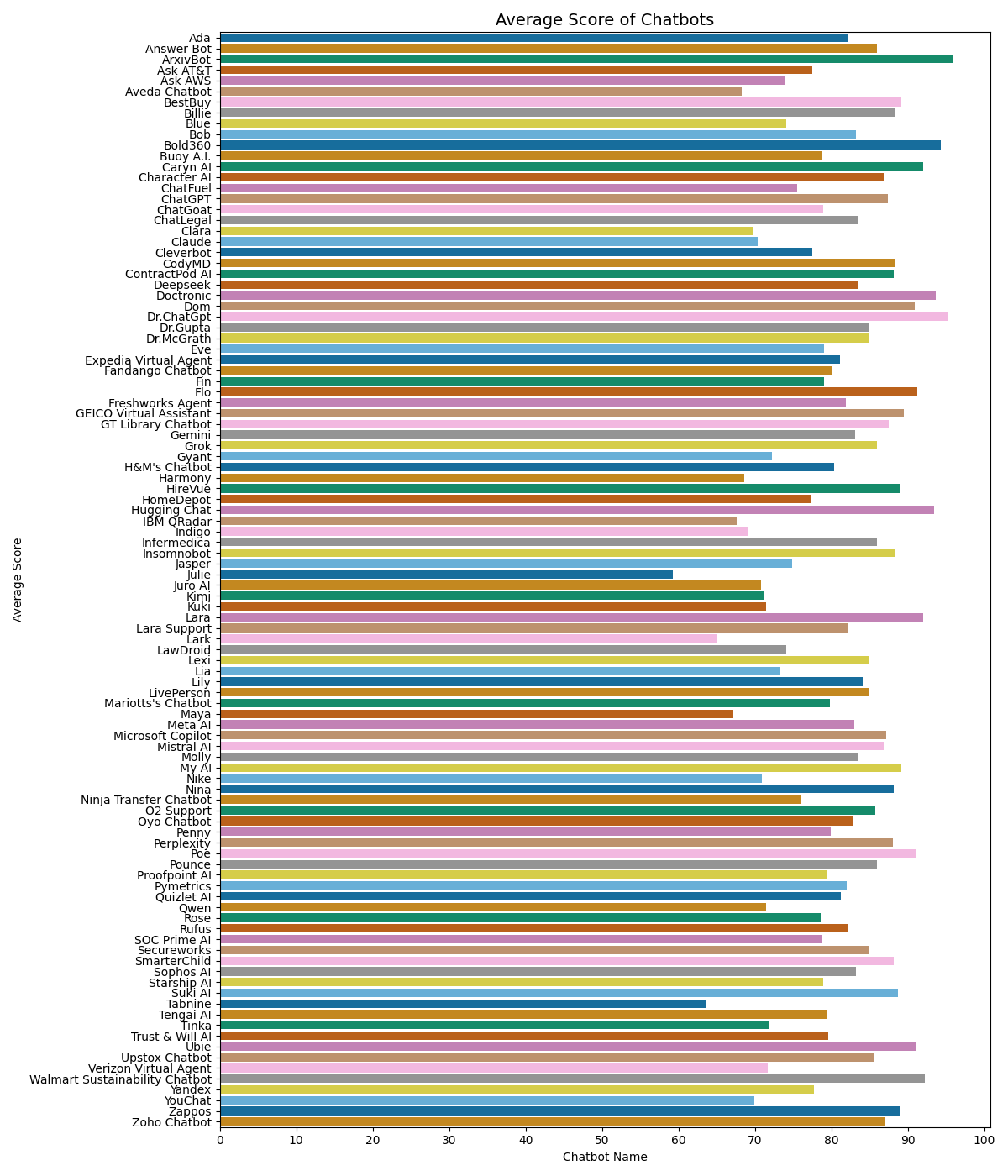}
  \caption{Bar Graph: Average scores across Google Lighthouse and PageSpeed Insights}
  \label{fig:avg_score}
  \Description{
A horizontal grouped bar chart showing average scores across multiple evaluation metrics for over 100 chatbot interfaces. Each row represents a chatbot, and the bars correspond to different scoring dimensions such as performance, accessibility, SEO, and best practices. Most chatbots score between 60 and 90, with visible variation in performance across individual metrics. The chart enables comparison of chatbot quality at a glance.
}
\end{figure*}

\nocite{saka2024evaluating}
\nocite{kishnani2023assessing}
\nocite{tazi2023accessibility}

\end{document}